\documentclass[twocolumn,trackchanges]{aastex631}

\usepackage{academicons}
\usepackage{color}
\definecolor{orcidlogocol}{HTML}{A6CE39}

\usepackage{inputenc}
\inputencoding{utf8}
\DeclareUnicodeCharacter{02BC}{\textquotesingle}
\usepackage{hyperref}
\usepackage{amsmath}

\usepackage{makecell}
\usepackage{enumerate}
\usepackage{graphicx}
\usepackage{float}
\usepackage{subfigure}
\usepackage{amsmath}
\usepackage{CJKulem}
\usepackage{threeparttable}
\usepackage{footnote}
\usepackage{appendix}
\begin{document}
\shortauthors{Hu et al.}

\title{A comprehensive comparison of spin-up and spin-down episodes of 4U 1538-522 observed with \textit{NuSTAR}}

\correspondingauthor{L. Ji}
\email{jilong@mail.sysu.edu.cn}

\author[0009-0009-7560-9084]{Y.F. Hu}
\affiliation{School of Physics and Astronomy, Sun Yat-Sen University, Zhuhai, 519082, People’s Republic of China}
\author[0000-0001-9599-7285]{L. Ji}
\affiliation{School of Physics and Astronomy, Sun Yat-Sen University, Zhuhai, 519082, People’s Republic of China}
\author[0000-0003-0454-7890]{C. Yu}
\affiliation{School of Physics and Astronomy, Sun Yat-Sen University, Zhuhai, 519082, People’s Republic of China}

\begin{abstract}
4U 1538-522 is a persistent high mass X-ray binary which exhibits secular spin evolution. In 2019, it underwent a torque reversal from spinning up to spinning down. We performed an extensive study using four {\it NuSATR} observations to compare temporal and spectral properties during different states. We observed no abrupt change in luminosity associated with the torque reversal. In addition, the pulse profile, the spectral shape and the power spectrum remained unchanged before and after the torque reversal. The orbital and super-orbital modulation profiles also showed no significant changes. We discuss possible mechanisms for the torque reversal and conclude that it is unlikely to be caused by interactions between the accretion disk and the magnetosphere. Instead, the transition of accretion modes in spherical accretion may be a plausible explanation.

\end{abstract}

\section{Introduction} \label{Introduction}
In neutron star X-ray binaries, mass and angular momentum transfer occur when compact objects accrete material from a low-mass or high-mass companion star via the Roche lobe or stellar winds \citep{Davidson1973ApJ...179..585D,Iben1991ApJS...76...55I}. 
During accretion processes, both the luminosity and the rotational evolution are influenced by interactions between the magnetosphere and the accretion flow \citep{Lai2014EPJWC..6401001L}. 
They lead to the accretion torque exerted on the magnetized neutron star, occasionally resulting in transitions between spin-up and spin-down phases, known as ``torque reversals". 
These transitions can occur over timescales ranging from weeks to months or even years \citep{Beri2018MNRAS.475..999B}.
They have been discovered in many accreting pulsars, such as Cen X-3, Vela X-1, OAO 1657-415, GX 1+4, 4U 1626-67, and 4U 1907+09 \citep{Bildsten1997ApJS..113..367B,Chakrabarty1997ApJ...481L.101C,Yi1997ApJ...481L..51Y,Fritz2006A&A...458..885F,Inam2009MNRAS.395.1015I,Camero-Arranz2010ApJ...708.1500C,Camero-Arranz2012A&A...546A..40C,Beri2014MNRAS.439.1940B,Liao2022MNRAS.510.1765L}.
If the specific angular momentum of accreting matter is sufficiently high, an accretion disk will form outside the magnetosphere, otherwise accretion can proceed quasi-spherically \citep{Burnard1983, Shakura2012}.
In the disk-accretion scenario,
the accretion torque is expected to correlate with the mass accretion rate onto the neutron star and therefore with the X-ray luminosity \citep[e.g.,][]{Ghosh1979ApJ...234..296G, Wang1987A&A...183..257W, Wang1995,Bozzo2009A&A...493..809B}.
However, in many cases these two quantities are barely correlated or even anti-correlated, e.g., in GX 1+4 \citep{Chakrabarty1997ApJ...481L.101C}. 
In the literature, there are also updated models proposed to explain the torque reversals according to different theoretical assumptions, such as the retrograde or warping disk \citep{Makishima1988Natur.333..746M,Nelson1997ApJ...488L.117N,van1998ApJ...499L..27V,Wijers1999MNRAS.308..207W,Beri2015MNRAS.451..508B}, inner disk transitions between the optically thin and thick regimes \citep{Yi1997ApJ...481L..51Y,Vaughan1997astro.ph..7105V,Yi1999ApJ...516L..87Y,Dai2006A&A...451..581D}, oblique rotators \citep{Perna2006ApJ...639..363P,Benli2020MNRAS.495.3531B}, transitions between the weak propeller and spin-up phases \citep{Ertan2017MNRAS.466..175E,Ertan2018MNRAS.479L..12E,Ertan2021MNRAS.500.2928E,Gen2022A&A...658A..13G}. 
However, we note that all these models have their own limitations and the mechanism of torque reversals remains unknown.

On the other hand, the quasi-spherical accretion model has been introduced to explain the long-term spin period evolution of wind-accreting systems \citep{Bozzo2008, Shakura2012, González2012A&A...537A..66G,Gonz2018MNRAS.475.2809G,2015MNRAS.446.1013P}.
When the accretion rate is less than  $\sim 4\times10^{16}$ $\rm g/s$, the accreting matter subsonically settles down onto an extended quasi-static shell outside the magnetosphere, which mediates the transfer of angular momentum between the neutron star and the accreting matter \citep{Shakura2012}.
In this case, both spin-up and spin-down are possible even with prograde accreting matter.
At a higher accretion rate, due to rapid Compton cooling, the accretion becomes highly non-stationary and a free-fall gap above the magnetosphere appears \citep{Shakura2012}.

4U 1538-522 is a wind-fed persistent X-ray binary formed by a massive (17 $M_{\odot}$) B0Iab supergiant \citep{Reynolds1992MNRAS.256..631R} and a pulsating neutron star with a spin period of $P\sim$526\,s \citep{Varun2019MNRAS.484L...1V}.
Its orbital period is $\sim$3.7\,d with a eclipse phase of $\sim$0.6\,d, and its inclination is $\sim67^{\circ}$
\citep{Becker1977ApJ...216L..11B,Makishima1987ApJ...314..619M,Falanga2015A&A...577A.130F}. 
The distance to the source is estimated at $\sim$6.6 kpc measured by \emph{Gaia} \citep{Bailer2018AJ....156...58B}. 
Some studies suggest a circular orbit \citep{Makishima1987ApJ...314..619M,Baykal2006A&A...453.1037B}, while others propose an elliptical orbit with an eccentricity of 0.17-0.18 \citep{Clark2000ApJ...542L.131C,Mukherjee2006JApA...27..411M}. 
Since its discovery by \emph{Uhuru} \citep{Giacconi1974ApJS...27...37G}, three torque reversals in 4U 1538-522 have been observed between 1976 and 2022 \citep[for details, see Figure 4 in][]{Sharma2023MNRAS.522.5608S}.  
It exhibited a long term spin-down trend in 1976-1988 at a mean rate of $\dot{P} = 4.2(2) \times 10^{-9}$ s s$^{-1}$ \citep{Becker1977,Davison1977a, Cusumano1989, Nagase1989}, then followed by a spin-up phase from 1991 April to 1999 May at a rate of $-8.3(1) \times 10^{-9}$ s s$^{-1}$ \citep{Hemphill2013}.
Thus the first torque reversal was inferred to occur in 1988 or 1989 \citep{Rubin1997ApJ...488..413R, Sharma2023MNRAS.522.5608S}. 
Subsequent \emph{Fermi}/GBM observations showed a steady spin-down at a mean rate of $4.9(1) \times 10^{-9}$ s s$^{-1}$ from 2008 to 2018.
Considering also the previous spin-up trend, \citet{Hemphill2013} suggested that the second torque reversal happened in 2009. 
After that, \emph{Fermi}/GBM detected another steady spin-up phase, indicating the third torque reversal occurring in 2019 May \citep{Tamang2024MNRAS.527.3164T}.
After 2020 September, the source does not present a secular spin evolution, and instead exhibits a sinusoidal spin variation \citep{Sharma2023MNRAS.522.5608S}.

In this paper, we aim to perform an extensive study of 4U 1538-22 around its third torque reversal, with particular emphasis on comparing temporal and spectral properties at different spin evolution epochs.
The paper is organized as follows: Section \ref{Observations} briefly describes observations and data analysis.
Results are presented in Section \ref{Data analysis and Results}.
We discuss and summarize theoretical models in Section \ref{Discussion}.

\section{Observations}\label{Observations}
The \emph{Nuclear Spectroscopic Telescope Array (NuSTAR)} consists of two co-aligned grazing incidence X-ray telescopes (FPMA and FPMB), covering in a wide energy range of 3-79 keV \citep{Harrison2013ApJ...770..103H}. 
It provides an collective field of view of $10^\prime$, an angular resolution of $18^{\prime \prime}$ along with an energy resolution of 400 eV at 10 keV and 900 eV at 60 keV. 
\textit{NuSTAR} conducted four observations of 4U 1538-522 in August 2016, May 2019, and February 2021 with a total exposure of 124\,ks.
In this paper, we refer them as observations A, B, C and D.
We used the NUSTARDAS v2.1.2 and the CALDB version 20211020\footnote{\url{https://heasarc.gsfc.nasa.gov/docs/heasarc/caldb/caldb_supported_missions.html}}, provided with the HEASOFT v6.31.1\footnote{\url{https://heasarc.gsfc.nasa.gov/docs/software/lheasoft/}}, to perform the standard data reduction procedure recommended by the official users' guide\footnote{\url{https://heasarc.gsfc.nasa.gov/docs/nustar/analysis/nustar_swguide.pdf}}. 
We screened raw event files using the {\sc NUPIPELINE} task, and extracted lightcurves and spectra using the {\sc NUPRODUCTS} task.
During the timing analysis, the barycentric correction was performed according to the source's coordinate RA=$235.59734^{\circ}$and Dec=$-52.38599^{\circ}$. 
When extracting spectra, we considered a source-centered circular region with a radius of $\sim 78^{\prime \prime}$ for both FMPA and FMPB, while the background was estimated from a region close to the source. 
As reported by \cite{Hemphill2019ApJ...873...62H} and \cite{Sharma2023MNRAS.522.5608S}, this source shows eclipsing behaviors, and we only focused on its out-of-eclipse phases.
During the spectral analysis, the energy band we studied was 3-50\,keV, above which the instrumental background is dominating. 
In this paper, all uncertainties quoted correspond to a 90\% confidence level.

\begin{table*}
\begin{center}
	\caption{A summary of \emph{NuSTAR} observations of 4U 1538-522.}
	\label{obslog}
	\begin{tabular*}{\textwidth}{@{\extracolsep{\fill}} cccccc}
	\hline
		Observation & Observation ID & Observation Date & MJD & Exposure & Orbital phase$^{\rm a}$ \\
		& & yyyy-mm-dd & (d) & (ks) & (s) \\ 
		\hline
		A & 30201028002 & 2016-08-11 & 57611.81  & 43.83 & 0.823-1.086 \\[0.5ex]
		B & 30401025002 & 2019-05-02 & 58605.94  & 36.87 & 0.486–0.727 \\[0.5ex]
		C & 30602024002 & 2021-02-16 & 59261.32  & 21.81 & 0.272–0.409 \\[0.5ex]
		D & 30602024004 & 2021-02-22 & 59267.03  & 21.56 & 0.803–0.948 \\[0.5ex]
	\hline
    \multicolumn{6}{l}{$^{\rm a}$ 
    The orbital ephemeris was adopted from \citet{Hemphill2019ApJ...873...62H}. 
    }
	\end{tabular*}
\end{center}
\end{table*}

\section{Data analysis and Results} \label{Data analysis and Results}
\subsection{Long-term spin history and orbital modulation profiles}
\begin{figure*}
\centering
\includegraphics[width=0.8\textwidth]{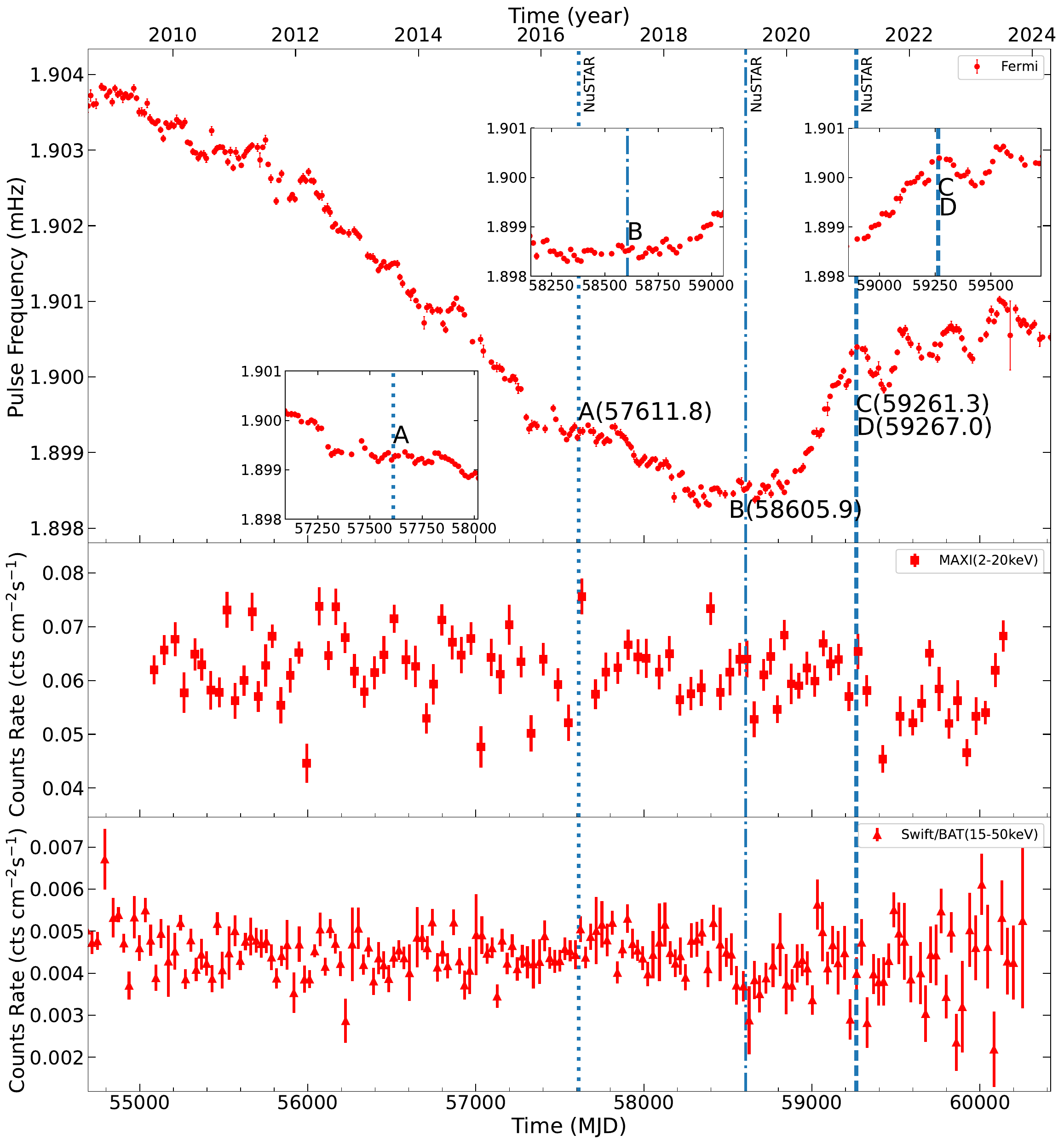}
\caption{
The pulse frequency observed with {\it Fermi}/GBM (upper), and long-term lightcurves of {\it MAXI} (2-20\,keV; middle) and {\it Swift}/BAT (15-50\,keV; bottom) of 4U 1538-522. 
For clarity, we re-binned the daily lightcurves with a time resolution of 30\,days. 
The vertical dashed lines indicate {\it NuSTAR} observations. 
The zoomed-in insets in the upper panel show the frequency evolution trend near each {\it NuSTAR} observation.
}
\label{counts} 
\end{figure*}

The spin frequency of 4U 1538-522 shows a secular evolution and switches between spin-up and spin-down episodes (Figure~\ref{counts}).
As already reported by \citet{Sharma2023MNRAS.522.5608S} and \citet{Tamang2024MNRAS.527.3164T}, the spin-down trend continues for about 10 years from the beginning of {\it Fermi}/GMB monitoring until $\sim$ MJD\,58400.
We note that there are also short-term spin evolutions superimposed on this long-term trend.
After that, the spin frequency remains almost constant for about one year, which could be regarded as a period of a slow torque reversal connecting the spin-up and spin-down episodes.
After $\sim$ MJD 58800, a clear spinning-up is detected.
In addition, after $\sim$ MJD 59115, the source only presents weak sinusoidal frequency fluctuations instead of a long-term spin-up or spin-down trend.
Our four \textit{NuSTAR} observations were made during different stages of the spin evolution.
The first observation was during the spin-down phase, while the second occurred during the constant spin phase, corresponding to the situation when a torque reversal might happen.
The third and fourth were detected when the spin evolution shows oscillations.

On the other hand, we investigated the long-term lightcurves observed with {\it MAXI} and {\it Swift}/BAT in the energy ranges of 2-20\,keV and 15-50\,keV, and did not discover significant variations associated with the spin evolution. 
In general, the flux of this source is quite stable in the past ten years, although there appears to be a slight decrease after 2021 observed with {\it MAXI}.

We also studied whether there were changes in orbital modulation profiles associated with the spin evolution.
We assumed that the torque reversal was occurred at MJD 58620 adopted from the linear fitting by \citet{Tamang2024MNRAS.527.3164T}, although the exact point can not been identified because of the slow evolution.
Thus, in practice, we divided {\it Swift}/BAT orbital lightcurves\footnote{
We did not use daily averaged lightcurves since the orbital period of 4U 1538-522 is only about 3.7 days.} 
into three groups, corresponding to the spin-down (time $<$ MJD 58620), spin-up (MJD 58620 $<$ time $<$ MJD 59115) and spin-fluctuation (time $>$ MJD 59115) epochs.

\begin{figure}
\centering
\includegraphics[width=0.49\textwidth]{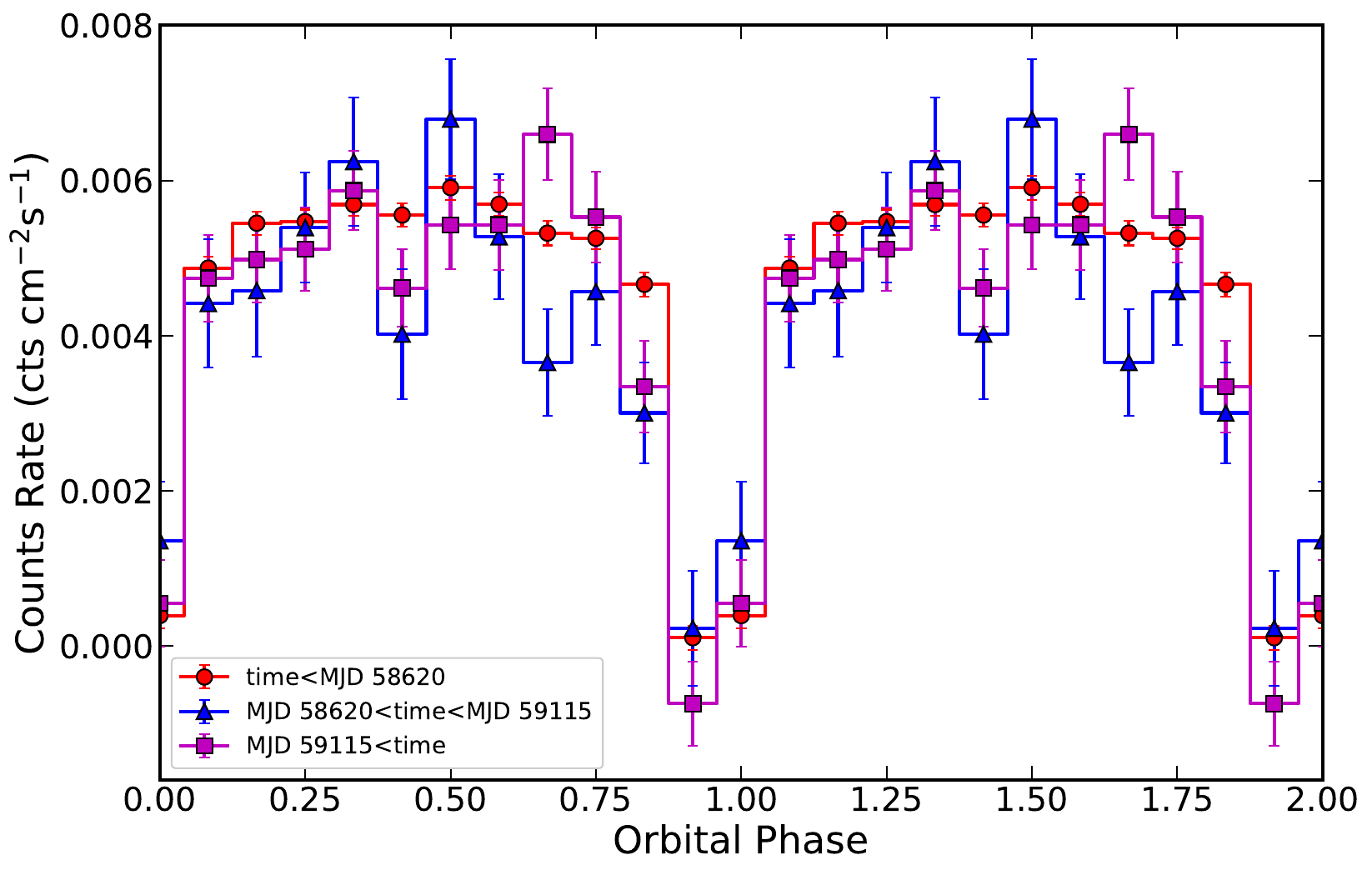}\vspace{-0.0cm}
\caption{Orbital modulation profiles of 4U 1538-522 observed with {\it Swift}/BAT, where red and blue points represent spin-down and spin-up epochs (i.e., time $<$ MJD 58620 and MJD 58620 $<$ time $<$ MJD 59115), and purple points represent the epoch showing spin fluctuations (time $>$ MJD 59115). 
Here we adopted the ephemeris from \citet{Hemphill2019ApJ...873...62H}.}
\label{orbital_profile}
\end{figure}

For each group, we folded the lightcurves using the ephemeris reported by \citet{Hemphill2019ApJ...873...62H}
, where the phase zero corresponds to the center of the eclipse.
As shown in Figure~\ref{orbital_profile}, the three orbital modulation profiles appear to be quite similar, with only some discrepancies around the phase 0.7.
We performed $\chi^2$-test to evaluate the variations between them, but only resulting in a marginal significance level of $\lesssim$2\,$\sigma$.
In addition, as reported by \citet{Corbet2021ApJ...906...13C}, 4U 1538-522 presents super-orbital modulation with a period of 14.913$\pm$0.0026 days.
We also explored potential changes in super-orbital modulation profiles, and the conclusion is similar to that of the orbital modulation, i.e., no discernible variations were observed.

\subsection{Spin pulse profiles}\label{Pulsar period and pulse profiles}
\begin{figure*}
\centering
\includegraphics[width=0.7\textwidth]{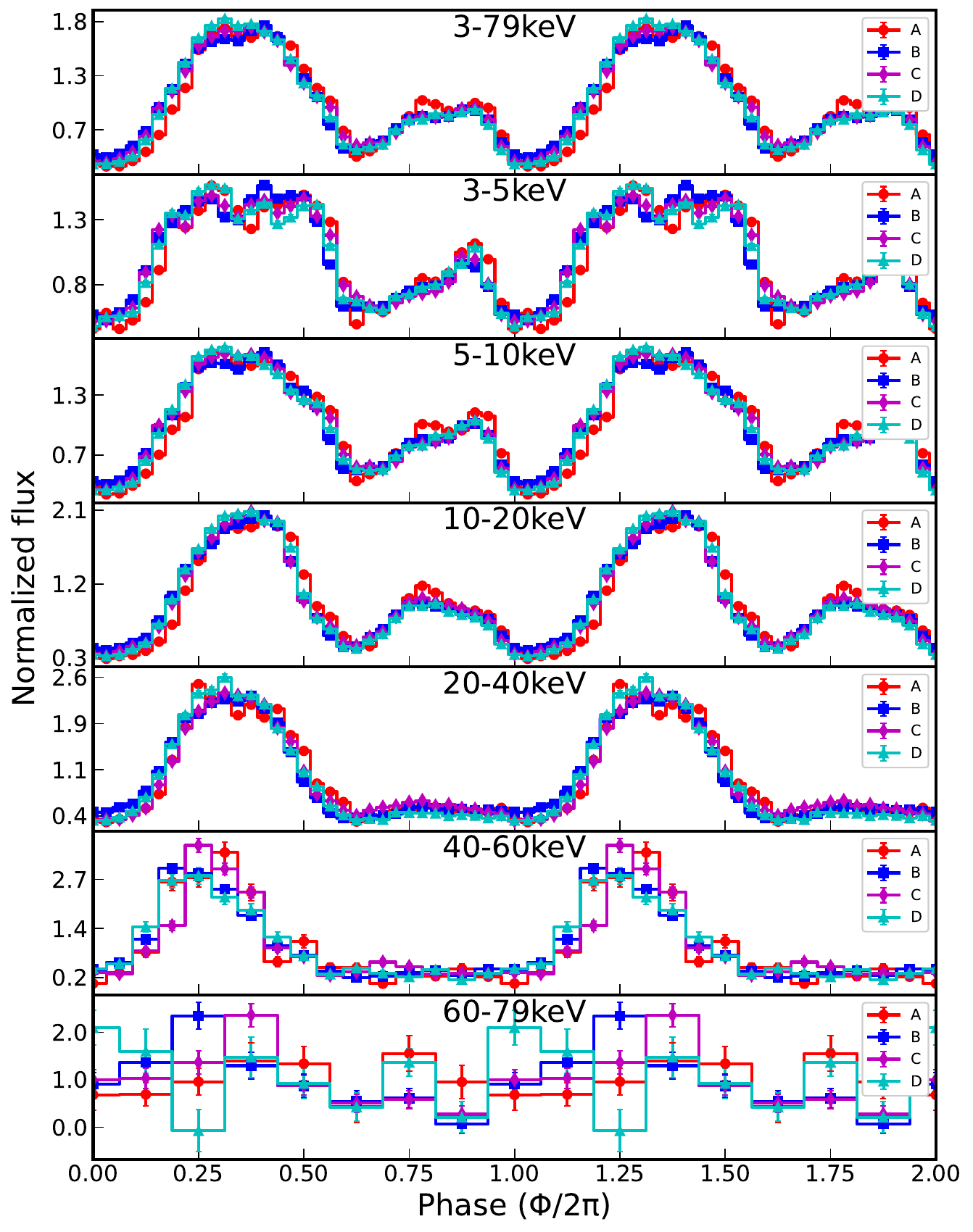}\vspace{0.0cm}
\caption{Energy-resolved pulse profiles of 4U 1538-522 in seven different energy bands (from top to bottom, 3-79\,keV, 3-5\,keV, 5-10\,keV, 10-20\,keV, 20-40\,keV, 40-60\,keV, and 60-79\,keV).
Different colors (red, blue, purple and cyan) represent 1-4 {\it NuSTAR} observations.
For the sake of comparison, all pulse profiles were normalized to their mean values.
}
\label{profile}
\end{figure*}

We compared broadband pulse profiles of these four observations.
In practice, we determined the spin period for each observation by using the epoch folding technique implemented by the ftool task {\sc efsearch} \citep{Leahy1983ApJ...266..160L}. 
The resulting spin periods are well in agreement with the results of Fermi/GBM \citep{Meegan2009ApJ...702..791M}.
We then generated background-subtracted lightcurves in seven different energy ranges (i.e., 3-5\,keV, 5-10\,keV, 10-20\,keV, 20-40\,keV, 40-60\,keV, 60-79\,keV and the total 3-79\,keV) for each observation
and folded pulse profiles by using the tool \texttt{efold} from the \textsc{XRONOS} package.
The results are shown in Figure~\ref{profile}.
For clarify, we aligned the pulse profiles using the cross-correlation method in Python module Scipy\footnote{\url{https://docs.scipy.org/doc/scipy/reference/generated/scipy.signal.correlate.html}}.
%
In general, pulse profiles of 4U 1538-522 are energy-dependent, which exhibit a double-peaked structure at low energies ($\lesssim$20\,keV), consisting of a broad main peak and and a weaker secondary peak.
At higher energies, the main peak of pulse profiles becomes narrower and sharper with the disappearance of the secondary peak.
Our results are consistent with previous reports \citep[e.g.,][]{Sharma2023MNRAS.522.5608S,Tamang2024MNRAS.527.3164T}.
We note that even though these four observations occurred at different stages of the spin evolution, their pulse profiles are remarkably similar.
This may suggest that there are no significant changes in the accretion structure near the surface of the neutron star.

\subsection{Power spectrum}\label{Power spectrum} 
\begin{figure}
\centering
\includegraphics[width=0.49\textwidth]{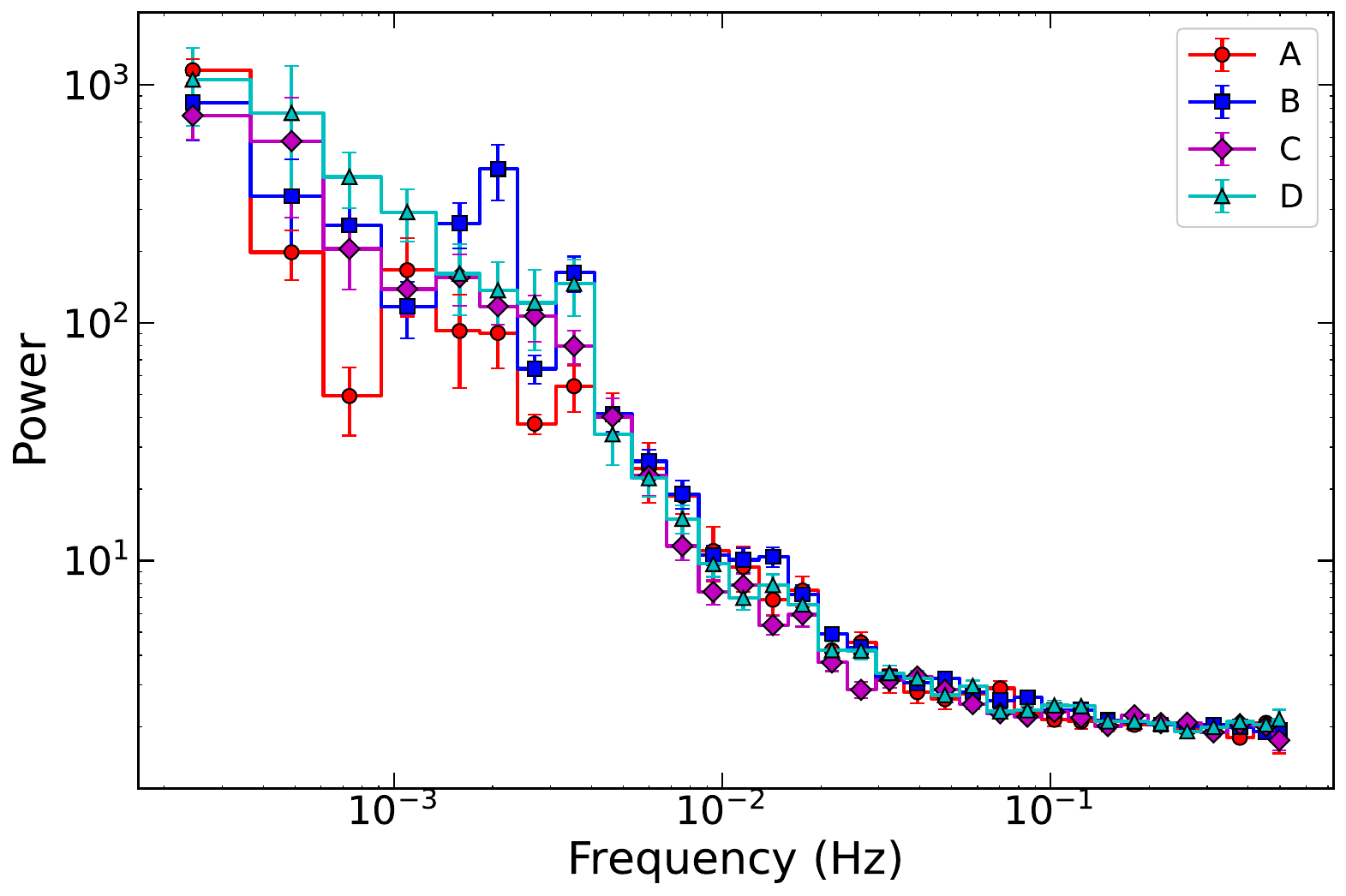}\vspace{-0.0cm}
\caption{Power density spectra of 4U 1538-522 obtained from four {\it NuSTAR} observations (A-D).
}
\label{powspectrum}
\end{figure}

For each {\it NuSTAR} observation, we calculated the power spectrum of the 3-79\,keV lightcurve using the task \texttt{powspec} from the \textsc{XRONOS} package.
Here we adopted the rms normalization and only considered the frequency range $2^{-13}-2^{-1}$\,Hz, which was dominated by the broadband red noise.
In addition, to study the non-periodic variability, pulsations were subtracted off from the lightcurve by using the averaged pulse profile as a template.
The results are shown in Figure~\ref{powspectrum}.
In general, the power spectra of these four observations are quite similar, suggesting that there are no significant changes in the accretion environment outside the magnetosphere between spin-up and spin-down episodes.
The primary discrepancies between them occur at around 1\,mHz and 2\,mHz, which are likely caused by pulse-to-pulse variations considering the spin period $\sim$526\,s of this source. 
This result is different from observations in 4U 1626-67, where  strong QPOs were present only in the spin-down era, and the dramatic change of power spectra was explained as switches between different accretion modes with different geometry of the accretion
flow \citep{Beri2014MNRAS.439.1940B,Jain2010MNRAS.403..920J}.

\begin{figure*}
  \centering

  \subfigure{\includegraphics[width=0.4\linewidth]{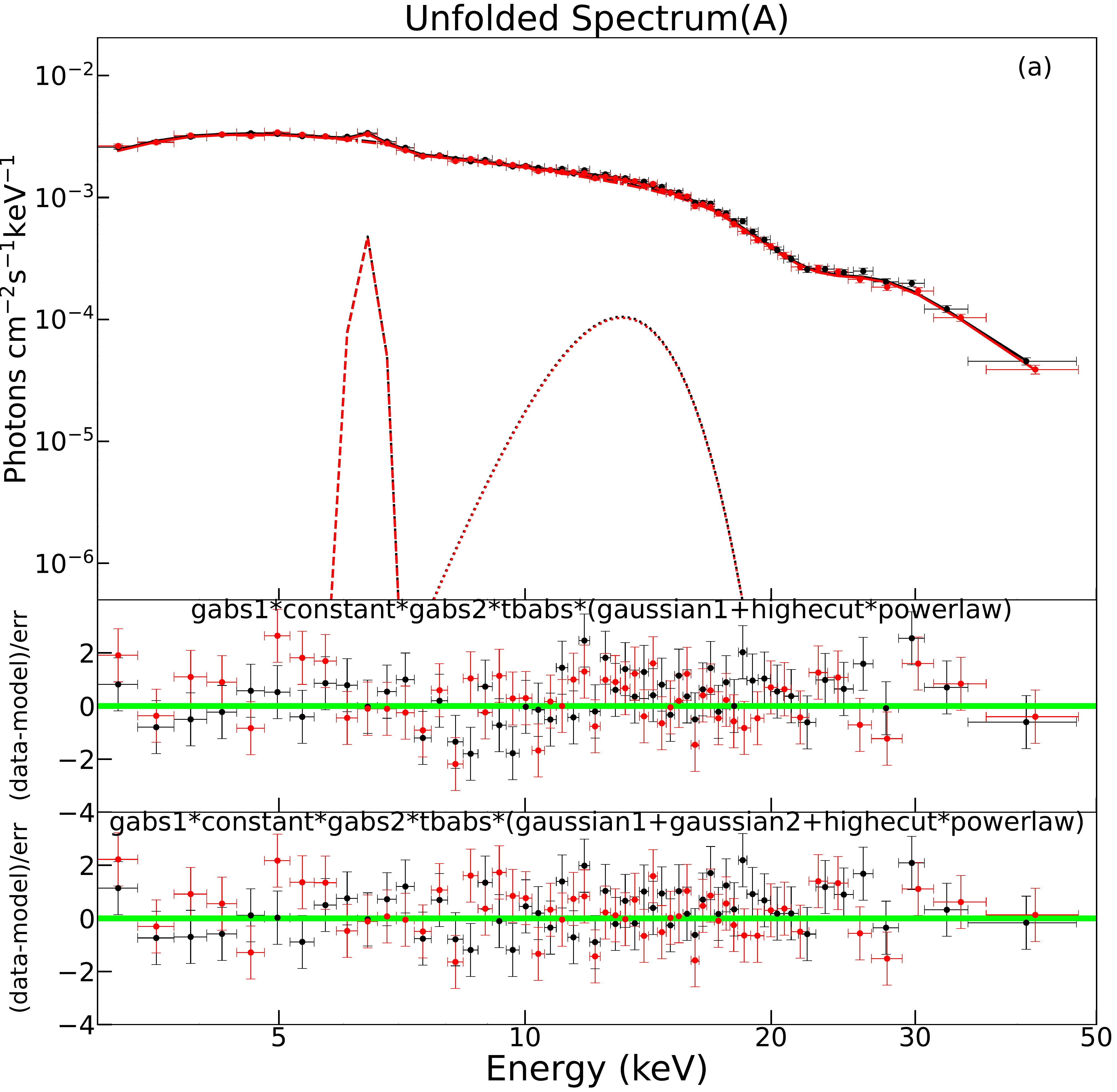}\label{fig:sub1}}
  \subfigure{\includegraphics[width=0.4\linewidth]{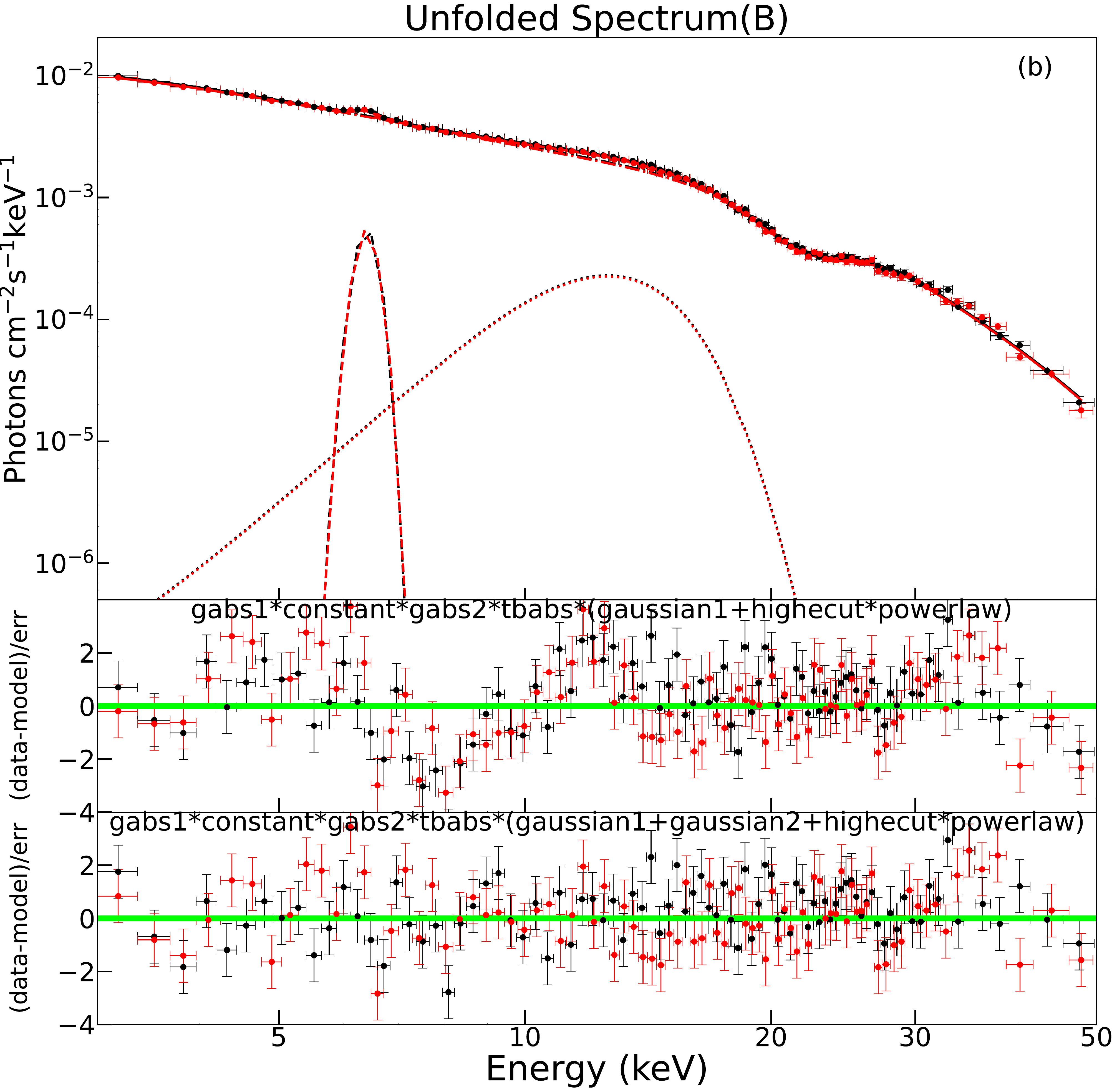}\label{fig:sub1}}
  \vspace{-0.3cm}
  \subfigure{\includegraphics[width=0.4\linewidth]{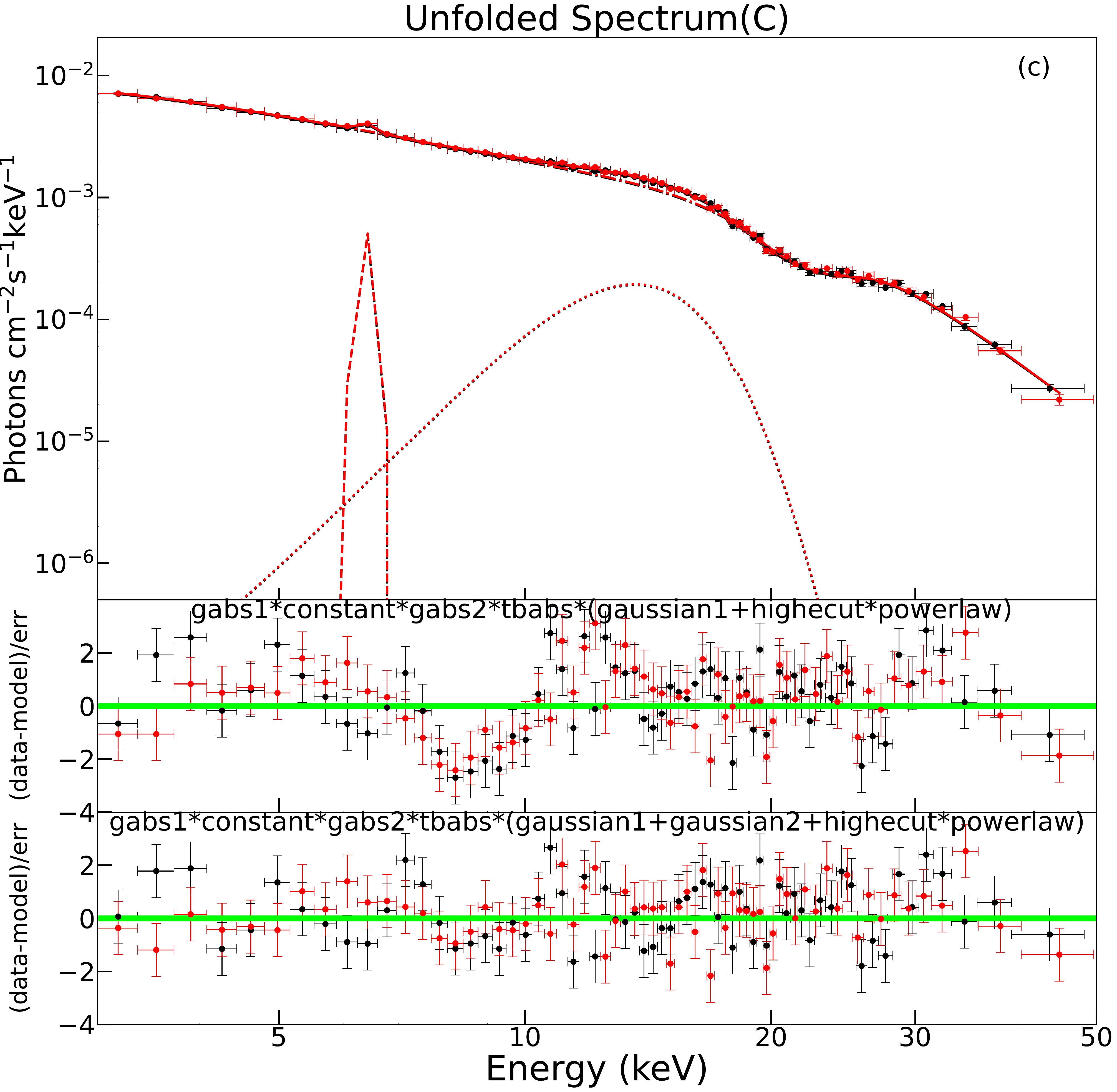}\label{fig:sub1}}
  \subfigure{\includegraphics[width=0.4\linewidth]{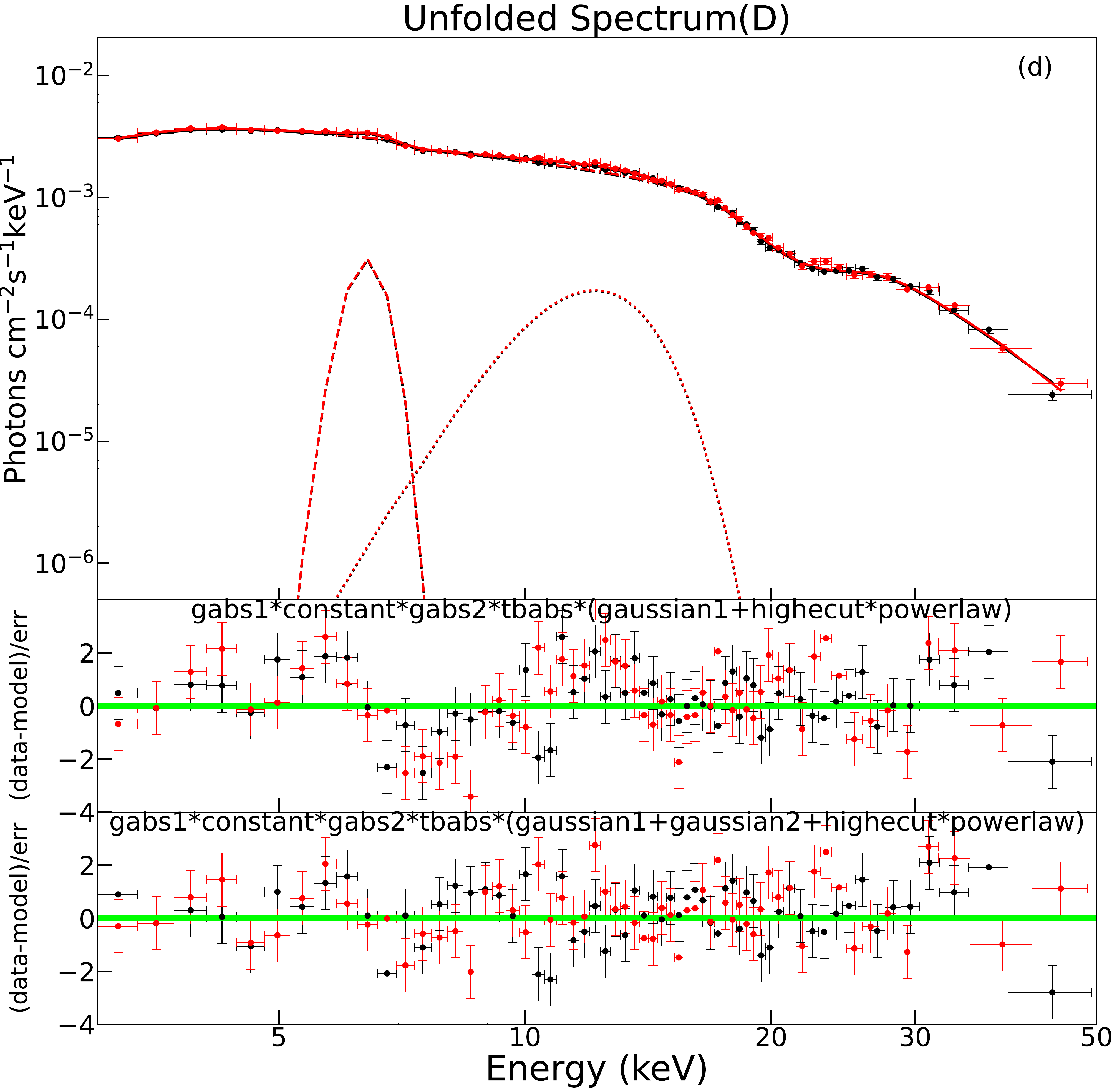}\label{fig:sub1}}

  \caption{Unfolded broadband spectra of 4U 1538-522 observed with {\it NuSTAR}/FPMA (red) and FPMB (black) using the model I \texttt{gabs1*constant*gabs2*TBabs(gaussian1 + gaussian2 + highecut*powerlaw)}. 
  Different spectral components are shown with dashed lines. 
  The middle and bottom panels indicate residuals without and with the additional \texttt{gaussian2} component to flatten the ``10 keV feature".
  }
  \label{spectrum1}
\end{figure*}

\begin{figure*}
  \centering

  \subfigure{\includegraphics[width=0.4\linewidth]{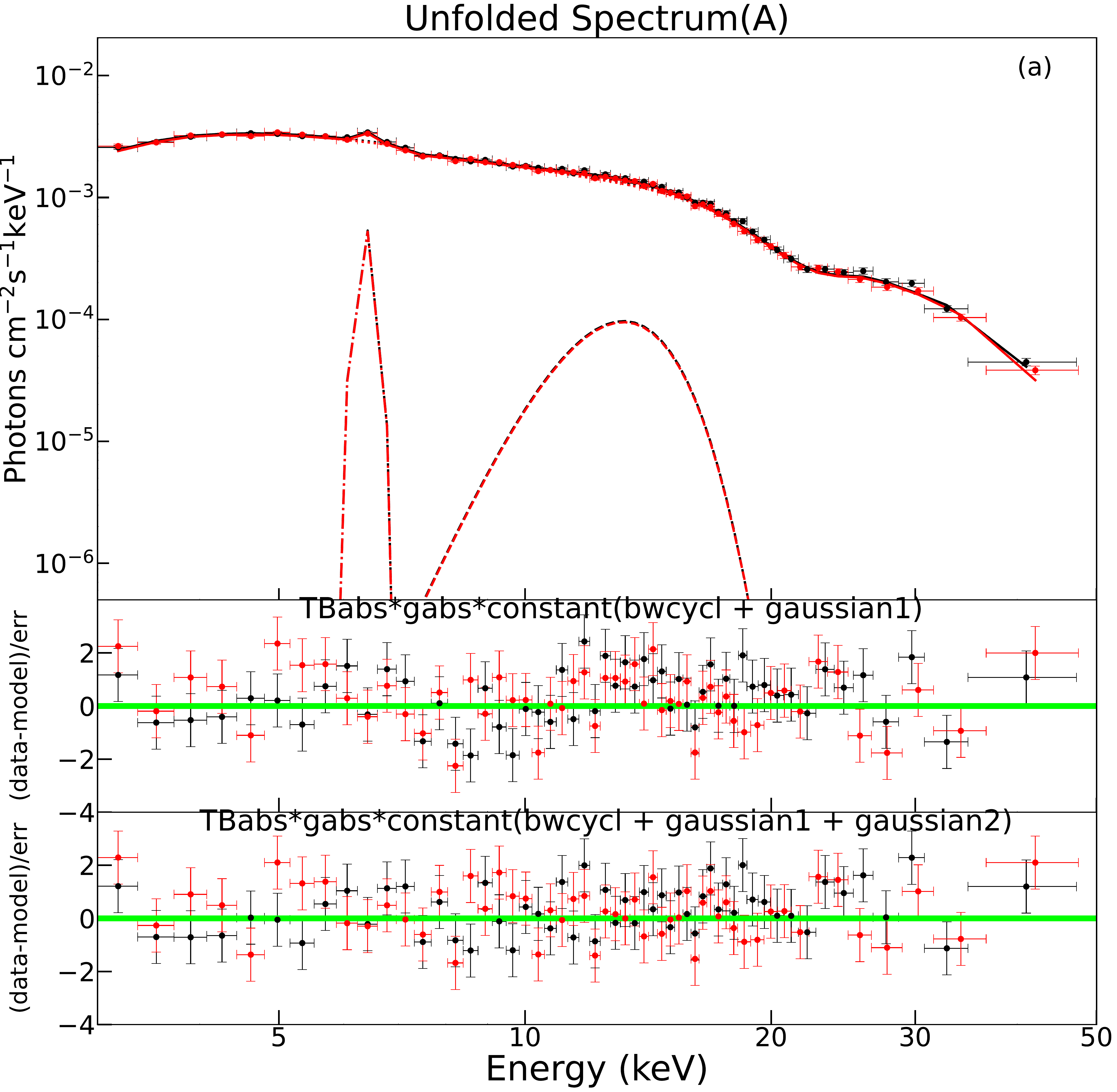}\label{fig:sub1}}
  \subfigure{\includegraphics[width=0.4\linewidth]{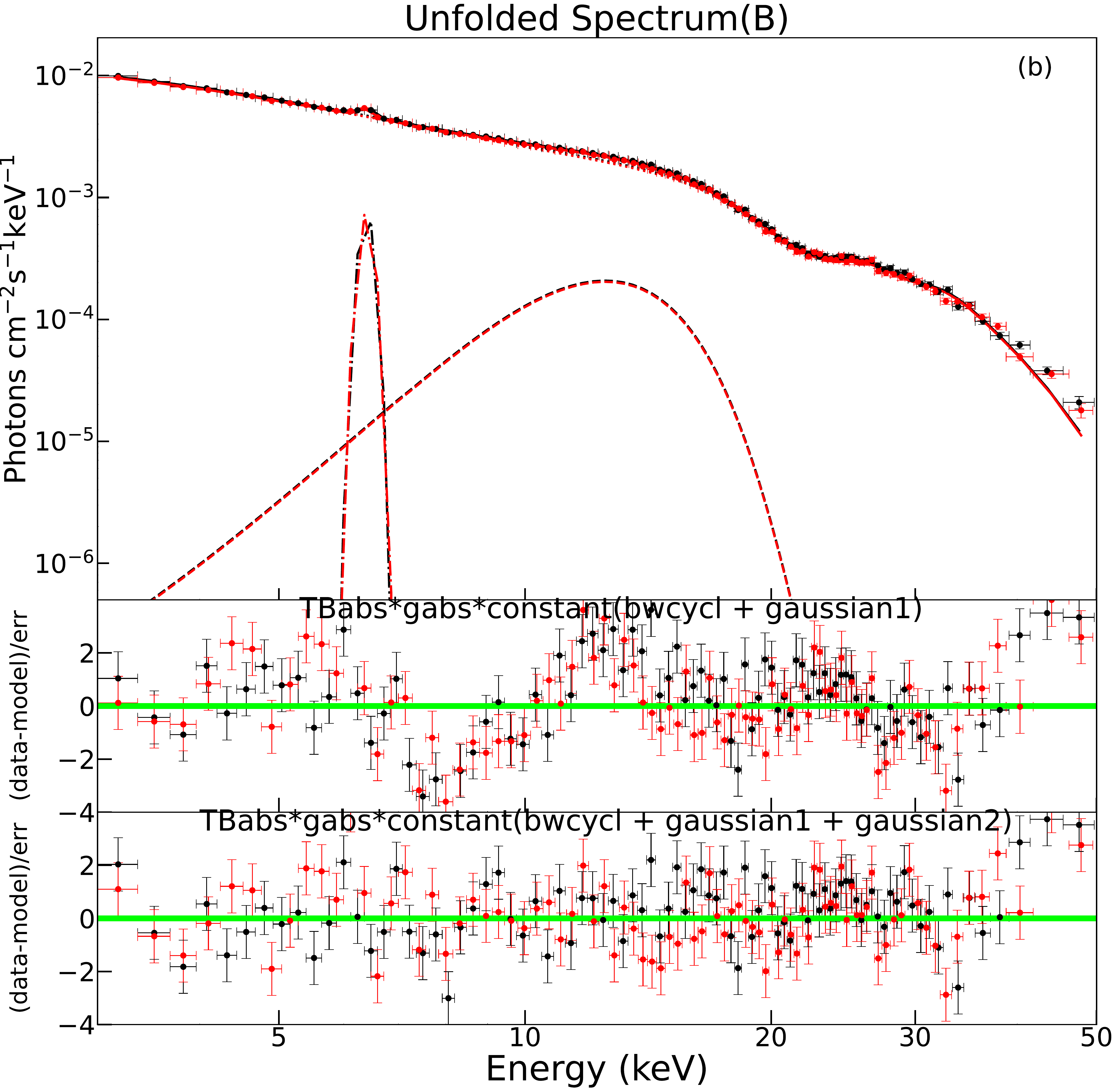}\label{fig:sub1}}
  \vspace{-0.3cm}
  \subfigure{\includegraphics[width=0.4\linewidth]{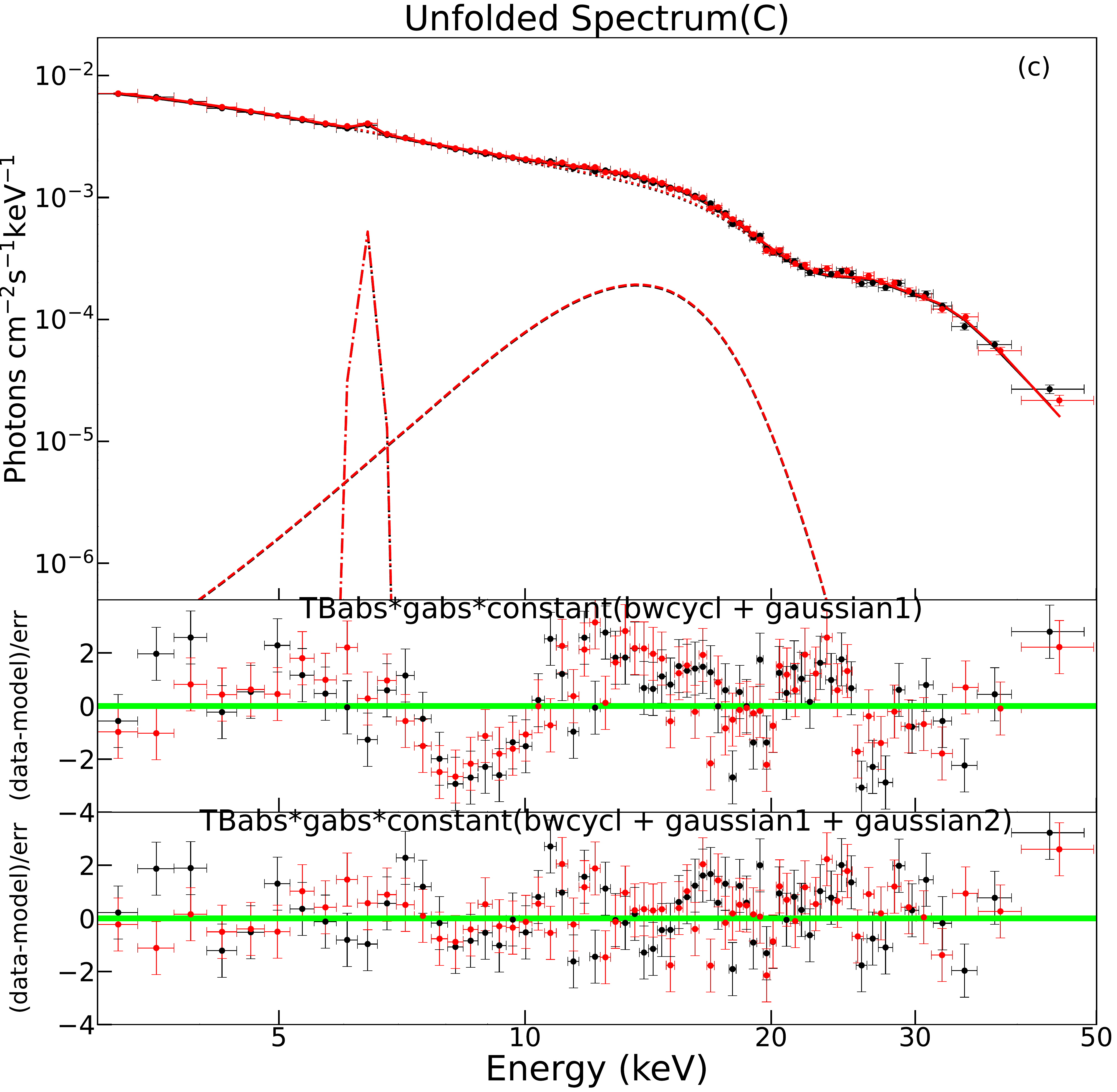}\label{fig:sub1}}
  \subfigure{\includegraphics[width=0.4\linewidth]{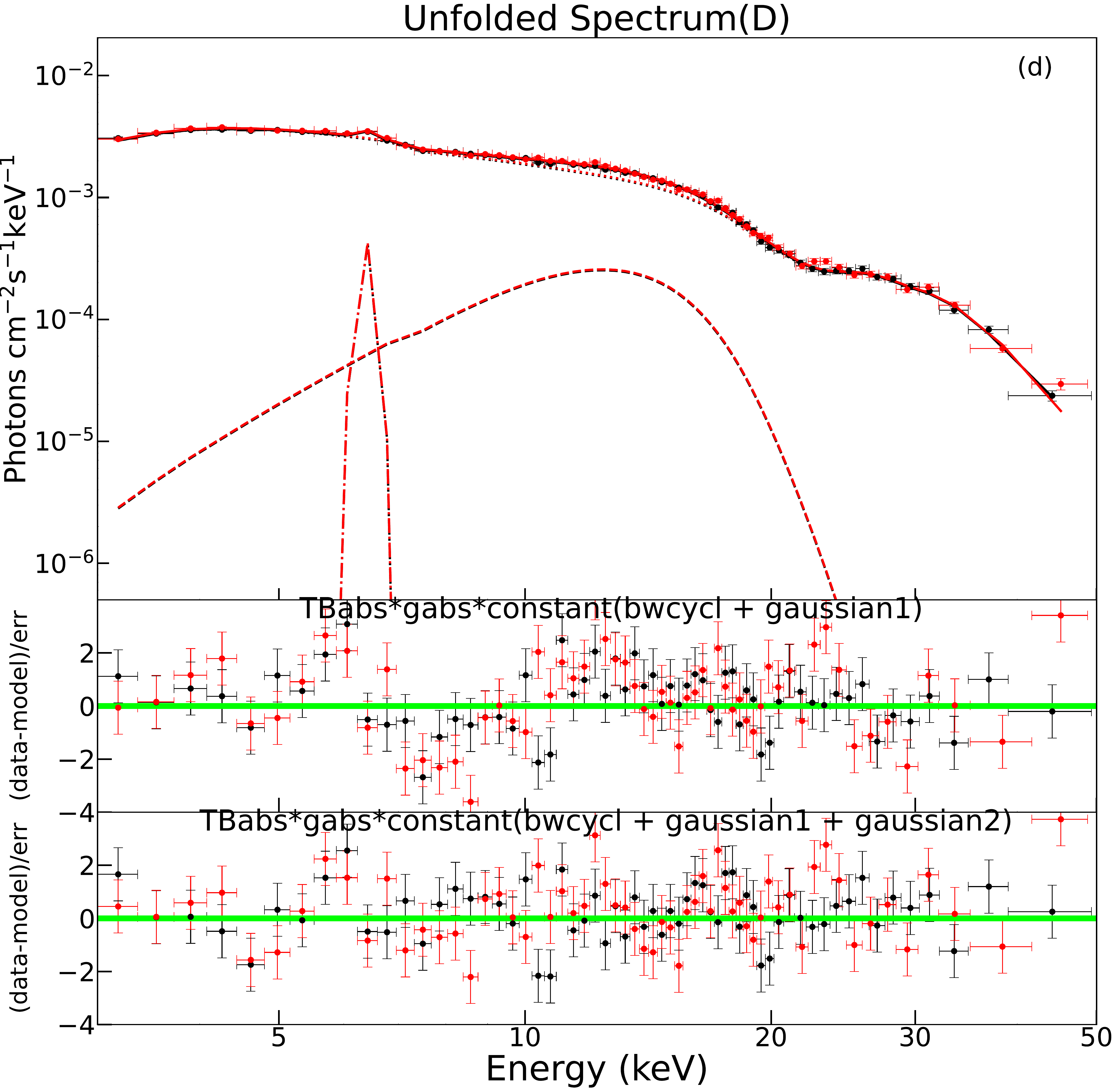}\label{fig:sub1}}

  \caption{
  Unfolded broadband spectra of 4U 1538-522 observed with {\it NuSTAR}/FPMA (red) and FPMB (black) using the model II \texttt{TBabs*gabs*constant(bwcycl + gaussian1 + gaussian2)}. 
  Different spectral components are shown with dashed lines. 
  The middle and bottom panels indicate residuals without and with the additional \texttt{gaussian2} component to flatten the ``10 keV feature".
  }
  \label{spectrum2}
\end{figure*}

\subsection{Spectroscopy}\label{Phase-averaged spectroscopy}
Following \citet{Hemphill2019ApJ...873...62H}, we started our spectral analysis with a phenomenological model, i.e., a product of a powerlaw model and a
multiplicative exponential factor (\texttt{highecut*powerlaw} in {\sc XSPEC}).
Since this model contains a discontinuity around the cutoff energy ($E_{\rm cut}$), we included a Gaussian absorption feature (\texttt{gabs1}) with the center energy fixed to $E_{\rm cut}$ in order to smooth the residuals \citep{Staubert2019}.
We also included a {\it tbabs} component to describe the photoelectric absorption from the interstellar medium \citep{Wilms2000}.
In addition, to account for the cross-calibration between FPMA and FPMB, we used a \texttt{constant} model that was fixed at 1 for FPMA and free for FPMB.
After initial fittings, we observed significant residuals around 6.4\,keV and 20\,keV.
According to previous studies \citep[e.g.,][]{Hemphill2019ApJ...873...62H,Tamang2024MNRAS.527.3164T,Sharma2023MNRAS.522.5608S}, they are likely caused by the Fe $K_\alpha$ fluorescence line and a cyclotron resonance scattering feature.
Therefore, we further included a Gaussian emission line (\texttt{gaussian1}) and an additional Gaussian absorption line (\texttt{gabs2}).
In general, such a model can provide a good description of the data, although around 10\,keV there is still a weak structure in residuals especially for observations B, C and D (see Figure~\ref{spectrum1}).
In literature, it is often referred to as the ``10 keV-feature" \citep{Coburn2002,Manikantan2023MNRAS.526....1M}.
Therefore, we further added an additional Gaussian emission component \texttt{gaussian2} to flatten the residuals\footnote{{ The ``10 keV-feature" in 4U 1538-22 was modeled as a emission feature by \citet{Hemphill2013ApJ...777...61H} or an absorption line/edge by \citet{Tamang2024MNRAS.527.3164T} and \citet{Sharma2023MNRAS.522.5608S}. We note that this model selection has little influence on the continuum parameters.}
}.
Thus the model we used was \texttt{gabs1*constant*gabs2 *tbabs*(gaussian1+gaussian2+highecut*powerlaw)} (model I).
The results are shown in Figure \ref{spectrum1} and Table \ref{table-highecut}.
Here we estimated the significance of the ``10 keV-feature" by using the \emph{Akaike Information Criterion} $\mathrm{AIC}=\chi^{2}+2k+(2k^{2}+2k)/(n-k-1)$, where $\chi^2$ is the chi-square value, $k$ is the number of free parameters and $n$ is the number of bins ($n-k$ is the number of degrees of freedom) \citep{K2020A&A...643A.128K}. 
The \emph{Chance Improvement Probability} of the additional component can be calculated as $\exp(-\Delta_{\mathrm{AIC}}/2)$.
Generally, the ``10 keV-feature" is more significant in observations B, C and D ($>5\,\sigma$), while in A it is marginally detected ($\sim3\sigma$).
Recently, the ``10 keV-feature" was systematically studied in different sources by \citet{Manikantan2023MNRAS.526....1M}. 
They only analyzed the observation A for 4U 1538-522 and did not detect significant fitting residuals around 10\,keV.
In addition, they found that the ``10 keV-feature" may only be present in some observations of a source, which is consistent with our results.
Other spectral parameters (e.g., the photon index and the cutoff energy) are relatively stable among four observations.
The cyclotron line remains around 22\,keV, without a clear long-term evolution.
The unabsorbed flux exhibits a $\sim$30\% variation across different observations, but it seems to be not correlated with the spin evolution.
For example, the Observation B has the largest flux while it does not present a spin-up trend.
Generally our results are similar to those obtained by previous studies \citep{Hemphill2019ApJ...873...62H, Tamang2024MNRAS.527.3164T, Sharma2023MNRAS.522.5608S} when using the same continuum model, although the  second absorption line around 28\,keV reported by \citet{Sharma2023MNRAS.522.5608S} was not detected in our residuals.

As a comparison, we also fitted the spectra with a  physical continuum model \texttt{bwcycl}, which was provided by  \citet{Becker2007ApJ...654..435B} and implemented by \cite{Ferrigno2009}.
This model considers thermal and bulk Comptonization of the seed photons provided by the thermal radiation, the cyclotron cooling and the bremsstrahlung in the accretion accretion column.
It has been successfully applied to many sources, such as 
Her X-1, 4U 1626-67, Swift J1845.7-0037, XTE J1858+034, Cen X-3 and 4U 1907+09 \citep{Wolff2016ApJ...831..194W, D'A2017MNRAS.470.2457D,Koliopanos2020arXiv200100723K,Malacaria2021ApJ...909..153M,Thalhammer2021A&A...656A.105T,Tobrej2023MNRAS.518.4861T}.
During the fitting, we assumed a distance $D$ of 6.6\,kpc \citep{Bailer2018AJ....156...58B} and characteristic values of a neutron star, i.e., a mass of $M_{*}=1.4\,M_{\odot}$ 
and a radius of $R_{*}=12$\,km.
%
The \texttt{bwcycl} model has six free parameters, which are the accretion rate $\dot{M}$, the magnetic field strength $B$, the accretion column radius $r_{\rm 0}$, the electron
temperature $T_{\rm e}$, the photon diffusion parameter $\xi$ and the Comptonization parameter $\delta$.
Similar to the model I, we also included the Fe $K_{\rm \alpha}$ line (\texttt{gaussian1}) and the cyclotron line (\texttt{gabs}) during the fits, as well as the \texttt{tbabs} and \texttt{constant} components.
In practice, since the cyclotron line energy ($E_{\rm cyc}$) is directly related to the local magnetic field as $B = \frac{1+z}{11.57} E_{\rm cyc} \times 10^{12}\,$G, where $z\approx0.3$ is the gravitational redshift near the surface of the neutron star, we linked these two parameters together.
In addition, the analytical calculation of the \texttt{bwcycl} model does not enforce the energy conservation, which means that the resulting energy-integrated luminosity does not strictly equal to the accretion luminosity ($GM_{*}\dot{M}/R_{*}$).
Following \citet{Wolff2016ApJ...831..194W}, we performed a iterative process.
We initially set $\dot{M}$ at the value calculated from the Model I by integrating the unabsorbed 3-50\,keV spectrum and assuming an isotropic radiation.
Then we estimated the accretion rate from the best-fitting parameters, which would be used to start the next iteration until convergence.
Similar to Model I, such a model could provide good fits, but still with evident residuals around 10\,keV especially for Observations B, C and D (see Figure~\ref{spectrum2}).
Thus, we again added an additional Gaussian component (\texttt{gaussian2}) to alleviate this feature and calculated its significance based on $\Delta AIC$.
Similar to the result of Model I, the ``10\,keV feature" is more significant in Observations B, C and D, which suggests that its presence might be model-independent.  
The final model is \texttt{TBabs*gabs*constant(bwcycl + gaussian1 + gaussian2)}(Model II), of which results are shown in Figure \ref{spectrum2} and Table \ref{table-bw}.
Generally we found an accretion column radius $r_{\rm 0}$ of several tens of meters and a plasma temperature $T_{\rm e}$ of $\sim$4\,keV, which are very similar to results in other accreting pulsars (e.g., Her X-1, 4U 1626-67, Cen X-3, Swift J1845.7-0037 and XTE J1858+034) \citep[][]{Wolff2016ApJ...831..194W,D'A2017MNRAS.470.2457D,Thalhammer2021A&A...656A.105T,Koliopanos2020arXiv200100723K, Malacaria2021ApJ...909..153M}.
The parameter $\delta$ is relatively small, suggesting that thermal Comptonization is the dominating process.
During some fits, the photon diffusion parameter $\xi$ was pegged at the hard limit, which is much larger than that in other sources.
This may suggest that the photon escaping through the column wall is more important in 4U 1538-222 \citep[see equation 104 in][]{Becker2007ApJ...654..435B}.
For these four observations, no dramatic changes of physical parameters $\xi$, $\delta$ and $T_{\rm e}$ are observed, and only the $r_{\rm 0}$ varies between 27-38 meters.
This indicates that the accretion structure near the polar caps does not undergo significant changes during different spin-up/down phases.

\begin{table*}
\setlength{\tabcolsep}{1.6pt}
\caption{Best-fitting model parameters of \textit{NuSTAR} observations using the model I: \texttt{gabs1*constant*gabs2*tbabs*(gaussian1+gaussian2+highecut*powerlaw)}
All the errors are quoted at 90 percent confidence level.}	
\label{table-highecut}
\begin{centering}
\begin{tabular}{ccccccc} 
\hline
Model & Parameters & Units & A & B & C & D \\
\hline
gabs1 & Sigma & keV & 0.61$_{-0.44}^{+3.88}$ & 0.23$_{-0.20}^{+0.24}$ & 0.01$_{-0.01}^{+0.01}$ & 0.02$_{-0.01}^{+0.01}$ \\

& Strength &  & 0.10$_{-0.07}^{+0.67}$ & 0.05$_{-0.03}^{+0.03}$ & 0.06$_{-0.06}^{+0.07}$ & 0.01$_{-0.01}^{+0.03}$ \\

gabs2 & LineE & keV & 21.73$_{-0.23}^{+0.42}$ & 21.52$_{-0.12}^{+0.14}$ & 21.32$_{-0.04}^{+0.03}$ & 21.51$_{-0.24}^{+0.18}$ \\

& Sigma & keV  & 3.27$_{-0.68}^{+0.19}$ & 3.21$_{-0.18}^{+0.14}$ & 3.33$_{-0.01}^{+0.01}$ & 3.19$_{-0.13}^{+0.16}$ \\

& Strength &  & 5.61$_{-2.57}^{+0.63}$ & 5.11$_{-0.46}^{+0.40}$ & 5.56$_{-0.05}^{+0.05}$ & 5.35$_{-0.38}^{+0.56}$ \\

$C_{\rm FPMB/FPMA}$ & factor &  & 0.98$_{-0.01}^{+0.01}$ & 0.98$_{-0.004}^{+0.004}$ & 1.01$_{-0.003}^{+0.003}$ & 1.02$_{-0.007}^{+0.007}$ \\

TBabs& nH & $10^{22}$ cm$^{-2}$ & 9.03$_{-0.32}^{+0.51}$ & 1.56$_{-0.20}^{+0.21}$ & 1.66$_{-0.14}^{+0.20}$ & 6.82$_{-0.33}^{+0.25}$ \\






gaussian2 & LineE & keV & 13.16$_{-0.57}^{+0.41}$ & 12.68$_{-0.41}^{+0.48}$ & 13.82$_{-0.25}^{+0.21}$ & 12.15$_{-0.35}^{+0.49}$ \\

& Sigma & keV & 1.68$_{-0.38}^{+0.36}$ & 2.64$_{-0.18}^{+0.13}$ & 2.72$_{-0.10}^{+0.14}$ & 1.85$_{-0.19}^{+0.28}$ \\

& Significance &  & $3.3\ \sigma$ & $11.0\ \sigma$ & $8.9\ \sigma$ & $7.6\ \sigma$ \\

highecut & cutoffE & keV & 18.52$_{-3.29}^{+1.46}$ & 18.05$_{-0.29}^{+0.22}$ & 17.90$_{-0.18}^{+0.16}$ & 17.0$_{-0.34}^{+0.51}$ \\

& foldE & keV & 10.55$_{-0.68}^{+1.28}$ & 10.77$_{-0.25}^{+0.45}$ & 10.78$_{-0.27}^{+0.33}$ & 10.11$_{-0.52}^{+0.40}$ \\

powerlaw & PhoIndex & & 1.23$_{-0.03}^{+0.03}$ & 1.30$_{-0.02}^{+0.03}$ & 1.29$_{-0.01}^{+0.01}$ & 1.10$_{-0.02}^{+0.02}$ \\
\hline
unabsorbed Flux (3-50\,keV) &   & {$\rm 10^{-10}erg/s/cm^2$} & 7.66$_{-0.31}^{+0.16}$ & 10.36$_{-0.1}^{+0.1}$ & 7.75$_{-0.03}^{+0.02}$ & 8.04$_{-0.1}^{+0.1}$ \\ 

$\chi^2_{\rm red}(\rm d.o.f.)$ &  & & 972 (964) & 1576 (1458) & 1205 (1181) & 1139 (1085) \\
\hline
\end{tabular}
\end{centering}
\end{table*}

\begin{table*}
\setlength{\tabcolsep}{1.6pt}
\caption{Best-fitting model parameters of {\it NuSTAR} observations using the model II: \texttt{TBabs*gabs*constant(bwcycl + gaussian1 + gaussian2)}.
All the errors are quoted at 90 percent confidence level.}	
\label{table-bw}
\begin{centering}
\begin{tabular}{ccccccc} 

\hline
Model & Parameters & Units & A & B & C & D \\
\hline

TBabs & nH & $10^{22}$ cm$^{-2}$ & 8.49$_{-0.57}^{+0.58}$ & 0.99$_{-0.20}^{+0.13}$ & 1.03$_{-0.26}^{+0.25}$ & 7.37$_{-0.61}^{+0.36}$ \\

gabs & LineE & keV & 22.48$_{-0.28}^{+0.19}$ & 22.37$_{-0.11}^{+0.10}$ & 22.18$_{-0.18}^{+0.18}$ & 22.21$_{-0.17}^{+0.23}$ \\

& Sigma & keV & 3.44$_{-0.20}^{+0.17}$ & 3.39$_{-0.11}^{+0.16}$ & 3.60$_{-0.22}^{+0.14}$ & 3.49$_{-0.23}^{+0.23}$ \\

& Strength & & 7.09$_{-0.45}^{+0.52}$ & 7.06$_{-0.30}^{+0.38}$ & 7.75$_{-0.65}^{+0.56}$ & 7.53$_{-0.54}^{+0.86}$ \\


$C_{\rm FPMB/FPMA}$ & factor & & 0.98$_{-0.01}^{+0.01}$ & 0.98$_{-0.003}^{+0.002}$ & 1.02$_{-0.01}^{+0.004}$ & 1.02$_{-0.01}^{+0.01}$ \\

bwcycl & Radius & km & 12.0(fixed) & 12.0(fixed) & 12.0(fixed) & 12.0(fixed) \\

& Mass & Solar & 1.4(fixed) & 1.4(fixed) & 1.4(fixed) & 1.4(fixed) \\

& $\xi$ & & 18.2$_{-3.09}^{+1.66}$ & 20.0$_{-0.55}^{+p}$ & 20.0$_{-0.54}^{+p}$ & 20.0$_{-1.31}^{+p}$ \\

& $\delta$ & & 0.14$_{-0.01}^{+0.03}$ & 0.13$_{-0.002}^{+0.003}$ & 0.13$_{-0.002}^{+0.004}$ & 0.12$_{-0.01}^{+0.01}$ \\

& \emph{B} (Tied to $E_{\rm CRSF}$) & $10^{22}\rm G$ & 2.53(fixed) & 2.51(fixed) & 2.49(fixed) & 2.50(fixed) \\

& $\dot{M}$ (Tied to luminosity) & $10^{16}\rm g/s$ & 2.0(fixed) & 2.8(fixed) & 2.1(fixed) & 2.2(fixed) \\

& $T_{\rm e}$ & keV & 4.07$_{-0.05}^{+0.04}$ & 4.01$_{-0.02}^{+0.02}$ & 3.97$_{-0.02}^{+0.03}$ & 4.07$_{-0.02}^{+0.05}$ \\

& $r_0$ & m & 27.01$_{-2.43}^{+1.28}$ & 38.48$_{-0.55}^{+0.29}$ & 29.34$_{-0.47}^{+0.24}$ & 32.26$_{-1.23}^{+0.60}$ \\

& $D$ & kpc & 6.6(fixed) & 6.6(fixed) & 6.6(fixed) & 6.6(fixed) \\







gaussian2 & LineE & keV & 13.21$_{-0.74}^{+1.48}$ & 12.59$_{-0.25}^{+0.52}$ & 14.00$_{-0.80}^{+0.72}$ & 12.47$_{-0.69}^{+1.10}$ \\

& Sigma & keV & 1.77$_{-0.26}^{+0.85}$ & 2.64$_{-0.17}^{+0.31}$ & 2.90$_{-0.20}^{+0.21}$ & 3.46$_{-0.32}^{+0.79}$ \\

& Significance &  & $3.9\ \sigma$ & $13.3\ \sigma$ & $11.7\ \sigma$ & $8.4\ \sigma$ \\

\hline

$\chi^2_{\rm red}(\rm d.o.f.)$ &  &  & 990(967) & 1642(1461) & 1226(1184) & 1146(1088) \\

\hline

\end{tabular}
\end{centering}
\end{table*}

\section{Discussion and Conclusion}\label{Discussion}
Torque reversal is a peculiar phenomenon occurring on a time scale ranging from days to years, which have been observed in many accreting pulsars, such as 4U 1626-67, GX 1+4 and OAO 1657-415 \citep{Yi1997ApJ...481L..51Y,Camero-Arranz2010ApJ...708.1500C}.
However, due to their significant differences in observations, there are no convincing physical models to explain all the behaviors.
4U 1538-522 has experienced three major torque reversals spanning between 1976 and 2022 \citep{Rubin1997ApJ...488..413R, Hemphill2013ApJ...777...61H, Sharma2023MNRAS.522.5608S}. 
In this paper, we performed an extensive study around the third torque reversal in 2019 using four \emph{NuSTAR} observations, focusing on comprehensive comparisons of its temporal and spectral properties, which will be useful to shed a light on its underlying mechanism.

Around the torque reversal, no evident variations were observed in the long-term lightcurves (Figure~\ref{counts}). According to the above analysis, the change in X-ray luminosity before and after the torque reversal is small as well as the spin-down rate is close to the spin-up rate, which is also reported in 4U 1626-67, GX 1+4, and OAO 1657-415 \citep{Yi1997ApJ...481L..51Y,Camero-Arranz2010ApJ...708.1500C,Liao2022MNRAS.510.1765L}. This suggests that this spin evolution is not a function of the accretion rate, which is therefore against the canonical torque model proposed by \citet{Ghosh1979ApJ...234..296G}.
%
On the other hand, \citet{Nelson1997ApJ...488L.117N} proposed a model to explain torque reversals with a nearly constant mass accretion rate by alternating between episodes of prograde and retrograde accretion disks.
This idea was partially modified by \citet{van1998ApJ...499L..27V}, which suggests that the retrograde disk might be caused by a tilted inner accretion disk triggered by the strong irradiation from the central source \citep{Pringle1996MNRAS.281..357P} or the tidal force from the companion \citep{Wijers1999MNRAS.308..207W}.
This model can successfully explain the anti-correlation between the torque and the luminosity in GX 1+4 and OAO 1657-415 \citep{Chakrabarty1997ApJ...481L.101C, Liao2022MNRAS.510.1765L,González2012A&A...537A..66G}.
However, in 4U 1538-522 the luminosity keeps stable and we did not find such a anti-correlation.  
In addition, there are several issues for this model, e.g., how to form a steady retrograde disk, why the characteristic time scales of spin-up and spin-down are so comparable in some sources \citep{Fritz2006A&A...458..885F}, and what the return mechanism is from the high tilt back to low tilt angle with multiple torque reversals \citep{Yi1999ApJ...516L..87Y}.

The invariant pulse profiles and luminosity in 4U 1538-522 before and after the torque reversal resemble the results observed in 4U 1907+09 \citep{Fritz2006A&A...458..885F}.
They suggested that this might be attributed to an oblique rotator configuration as proposed by \citet{Perna2006ApJ...639..363P}.
The main idea is that the region of disk-magnetosphere interactions is an ellipse in the equatorial plane.
Thus, it is plausible that the propeller effect (i.e., matter is expelled by the centrifugal force) works locally in certain regions, while the accretion is not inhibited in others.
However, we note that 4U 1538-522 is a slow rotator, which means that its co-rotation radius $R_{\rm co}=(\frac{GMP^2}{4\pi^2})^{1/3}\approx10^5$\,km is much larger than the Alfv\'en radius $R_{A}=(\frac{\mu^4}{GM\dot{M}^2})^{1/7} \approx5\times 10^3$\,km.
This renders all models \citep[e.g.,][]{Perna2006ApJ...639..363P, Benli2020MNRAS.495.3531B, Ertan2021MNRAS.500.2928E} that rely on the propeller effect impossible.

Another possibility of torque reversals might be related to the transition between a Shakura–Sunyaev disk and an advection-dominated accretion flow (ADAF) in the inner accretion disk, which was used to interpret behaviors in 4U 1626-67 and GX 1+4 \citep{Yi1997ApJ...481L..51Y,Yi1999ApJ...516L..87Y}.
The transition is expected to occur at a critical accretion rate of $\dot{M}$ $\sim 10^{35}-10^{36}$\,g/s, which is roughly compatible with our observations.
However, in 4U 1538-522 we did not find a spectral hardening associated with the torque reversal, which would be expected due to the Compton up-scattered by the hot plasma in the ADAF.
Additionally, variations in accretion modes, such as changes of the thickness of the accretion flow outside the magnetosphere, would significantly impact pulse profiles and power spectra.
However, unlike the case in 4U 1626-67 \citep{Beri2014MNRAS.439.1940B, Jain2010MNRAS.403..920J}, these effects were not detected in our observations, 
which suggests that this physical scenario seems unlikely.
However, we would like to caution that the argument that pulse profiles and power spectra in 4U 1538-522 are independent of the spin evolution still needs further testing in the future. 
This is because all current {\it NuSTAR} observations happened by chance during periods without significant spinning-up or spinning-down (see zoomed-in plots in Figure~\ref{counts}), although they are situated within different long-term trends.

All models above assume a spin-down torque resulting from interactions between the accretion disk and the magnetosphere, albeit with different considerations.
However, in observations we did not find signals (e.g., Quasi-periodic oscillations) of the appearance of an accretion disk.
Alternatively, considering the accretion in 4U 1538-522 is via a spherically symmetric stellar wind \citep{Mukherjee2006JApA...27..411M}, it is more likely that the accretion takes place quasi-spherically.
In this case, as suggested by \citet{Shakura2012}, an extended quasi-static shell may form outside the rotating magnetosphere if the matter is sub-sonical when the accretion rate is $\dot{M}  < \dot{M_{*}} \simeq 4\times10^{16}$\,g/s.
This critical value $\dot{M_{*}}$ is comparable with the accretion rate observed in 4U 1538-522.
At a higher accretion rate, due to the rapid Compton cooling, accretion becomes highly non-stationary and therefore a free-fall gap above the magnetosphere will appear.
In this scenario, the torque reversal can be explained by the transition between different accretion modes.
In both cases, the matter leaks from the magnetospheric cusp \citep[where the magnetic force line is branched;][]{Arons1976} down to the neutron star.
The shell does not provide a strong scattering for the radiation emitted from the neutron star surface because of the small Thomson depth.
Therefore, it is not surprising that the pulse profile and the spectrum remain stable during the torque reversal.
If the shell exists, the large-scale convective motions may provide an intrinsic variability on the order of millihertz \citep{Shakura2012}.
Within this frequency range, we did not detect QPOs in power spectra. 
However, we note that this may be attributed to the presence of strong red noise and the influence of periodic signals at several millihertz as well.

The differences significance between the orbital modulation profile presented in Figure. \ref{orbital_profile} $\lesssim$2\,$\sigma$ confidence level, revealing that the torque reversal is almost independent of the orbital modulation effect. 
\cite{Corbet2021ApJ...906...13C} also discovered small variation in the orbital modulation and superorbital modulation profiles during the spin-down phases of 4U 1538-522. 
Our result is also similar to the superorbital modulation profiles with slight change during different spin phases in 4U 1909+07 and IGR J16418-4532\citep{Islam2023ApJ...948...45I}. 

\section{acknowledgments}
This work has been supported by the National Natural Science Foundation of China (Grant Nos. 12173103, 11521303, 11733010, 11873103, 12373071,  U2038101, U1938103 and 11733009), and the International Partnership Program of Chinese Academy of Sciences (Grant No.113111KYSB20190020).
This work has also been supported by the National SKA Program of China (Grant No. 2022SKA 0120101), and the National Key R$\&$D Program of China (No. 2020YFC2201200), the science research grants from the China Manned Space Project (No. CMS-CSST-2021-B09, CMS-CSST-2021-B12 and CMS-CSST-2021-A10), and the opening fund of State Key Laboratory of Lunar and Planetary Sciences (Macau University of Science and Technology) (Macau FDCT Grant No. SKL-LPS(MUST)-2021-2023). 

\bibliography{sample631}{}
\bibliographystyle{aasjournal}

\end{document}